# LAND MOBILE SATELLITE CHANNEL MODEL FOR AUTONOMOUS GPS POSITIONING ESTIMATION

Jia-Chyi Wu, National Taiwan Ocean University, Keelung, Taiwan, ROC
Dah-Jing Jwo, National Taiwan Ocean University, Keelung, Taiwan, ROC
Chun-Yu Liu, National Taiwan Ocean University, Keelung, Taiwan, ROC

**ABSTRACT**

*The performance of GPS receivers with vector delay-locked tracking loop (VDLL) structure over a land mobile satellite (LMS) channel model for the simulation of multipath fading transmission is investigated. We apply vector delay-locked tracking loop receiver for GPS signal transmitting over LMS multipath fading situations. The adeptness of the vector tracking loop is to incessantly position when satellite GPS signal outage occurred. However, miscarriage tracking in one channel may overspread into the whole system originated the entire loss of locking on all satellites for vector tracking loop algorithm. To moderate this locking-loss situation, the VDLL system is carried with receiver autonomous integrity monitoring (RAIM) and fault detection and exclusion (FDE) algorithm to avoid signal unlocked incidence. The satellite signal to GPS receiver designs with two different Kalman filters (EKF: Extended Kalman Filter and UKF: Unscented Kalman Filter) are established to compare their performance over LMS multipath fading channel model simulations.*

## 1. INTRODUCTION

Signal processing over the land mobile satellite (LMS) channel model can be realized independently of the transmitted signal, since the proposed channel model comprises time variant reflectors approaching and lessening in dependency of the azimuth and elevation of the satellite. Therefore, we have applied the LMS channel model to better experiment GPS receiver performance over LMS fading channel situations. The Global Positioning System (GPS) is a satellite-based navigation system providing an operator access to proper and accurate positioning information everywhere in the world [Spilker et al, 1996]. The satellite transmits carrier signal to GPS receivers suffered signal degradation in varied channel conditions, from space through atmosphere to ground, especially due to the effects of multipath fading, a considerable familiarity of the propagation channel, the so-called land mobile satellite (LMS) channel, becomes necessary. The LMS channel is mainly impaired by the environment in the direct vicinity of the moving terminal. In fact, signal transmitted over an open environment with the line of sight (LOS) direct path may be completely blocked by solid obstructs (e.g., a building, trees, Lampposts) or may be comparatively shadowed by flora depending on the density of the vegetation and branches. Instead, multipath components are generated by diffraction, reflection, and scattering of the transmitted signal. In an urban area, a NLOS (none line of sight) signal transmission is experienced, the LOS signal is often foiled so that the first received path is attenuated and possibly delayed. Therefore, multipath fading error is the code tracking loop error caused by the LOS direct signals of the satellite and the reflected signals in the vicinity of the receiver. To investigate and improve the system performance of GPS signal transmitting over these multipath fading situations, we have applied a vector delay-locked tracking loop (VDLL) structure [Pany et al, 2006; Lashley et al, 2009] to GPS receiver with the support of Kalman filtering algorithms [Lashley et al, 2011; Zhu et al, 2010].

The vector tracking loop algorithm applied to GPS receiver can be continuously positioning when satellite signal outage occurred. However, a major drawback for vector tracking loop algorithm is tracking failure in one channel may spread all-over the entire system caused the catastrophe loss of locking on all satellites. To prevent signal unlocked situation, the VDLL system is delivered with receiver autonomous integrity monitoring (RAIM) and fault detection and exclusion (FDE) algorithm in this study. As we have known, to improve the GPS navigation receiver position and velocity determination with minimum mean-square error (MMSE) estimate of the system state vector, the Kalman filtering scheme has been widely applied. In the

Kalman filtering, system model with initial conditions and noise characteristics need to be specified as a priori [Jwo 2005, 8]. It is unrealistic for the filtering performance degradation situation improved by the availability of a precisely known model, since there are vacillations in the system models and noise description, and the assumptions on the statistics of disturbances are dislocated in some practical situations. We have explored the extended Kalman filter (EKF) and the unscented Kalman filter (UKF) embed separately into the GPS receiver to improve the positioning performance. The state transition and observation models need not be linear functions of the state but may instead be non-linear functions of differentiable type in both the EKF/UKF situations. There are many error effects are condensed due to the improving GPS receiver design. As the GPS signal is in harsh environments, the traditional receiver easily loses lock, resulting in reduced measurement accuracy. Simulations of the satellite signal to GPS receiver over the multipath fading noise is investigated in this study under LMS multipath channel conditions.

This paper is organized as follows; the LMS multipath channel model for GPS navigation process is given in the next section. The following section, simulation experiments on GPS navigation processing are carried out to evaluate the performance of the approach in comparison to those by the EKF/UKF through LMS fading channels. Analysis for various combinations of the above mentioned algorithms is studied and compared. Conclusions are stated in the final section.

## 2. KALMAN FILTERING GPS NAVIGATION RECEIVER

For the GPS signal transmission over LMS multipath fading situations, a vector delay-locked tracking loop structure is applied to GPS receiver with the support of Kalman filtering algorithms. The vector delay-lock loop (VDLL) algorithm [Peng et al, 2012] combines the code tracking and navigation solution into Kalman filter structure which include the extended Kalman filter (EKF) and the unscented Kalman filter (UKF) in this study [Jwo et al, 2007; Jwo et al, 2009]. The vector-based structure investigating the complete relation among different channels performs progressively better in the signal tracking. There are three major benefits of the VDLL: increased availability, realistic Doppler accuracy and less computational load. Performance study results show that the vector-based method outperforms the scalar tracking loops and commercial GPS receiver in the high dynamics and weak signal environment [Chen, 2010; Li et al, 2011]. Conversely, failure tracking overspreading from one channel into the whole system initiated the intact loss of locking on all satellites is a major drawback for vector tracking loop algorithm. In a former study [Wu et al, 2017], we may reduce the effect of overspread failure tracking by employing receiver autonomous integrity monitoring (RAIM) and fault detection and exclusion (FDE) algorithm to prevent signal unlocked situation in vector tracking loop. The RAIM algorithm and the Kalman filter structure are briefly reviewed for convenience.

### 2.1. Kalman Filter

Kalman filtering is a promised state estimation technique with minimum estimation error variance. The Kalman filter algorithm is with initial condition values, when new measurement becomes available with the progression of time, the estimation of states and the resultant error covariance would follow recursively and infinity. The Kalman filter can be represented as

Process model: $\quad \dot{\mathbf{x}} = \mathbf{F}\mathbf{x} + \mathbf{G}\mathbf{u}$ (1a)

Measurement model: $\quad \mathbf{z} = \mathbf{H}\mathbf{x} + \mathbf{v}$ (1b)

where the vectors **u** and **v** are both noise sequences. Eq. 1 is expressed as a discrete-time equivalent form leads to

$$\mathbf{x}_{k+1} = \mathbf{\Phi}_k \mathbf{x}_k + \mathbf{\omega}_k \quad (2a)$$

$$\mathbf{z}_k = \mathbf{H}_k \mathbf{x}_k + \mathbf{v}_k \quad (2b)$$

where the state vector $\mathbf{x}_k$, process noise vector $\mathbf{\omega}_k$, measurement vector $\mathbf{z}_k$, and measurement noise vector $\mathbf{v}_k$, are all in $\mathbf{R}^n$. In Eq. 2, the state vectors $\mathbf{x}_k$ and $\mathbf{\omega}_k$ are assumed to be zero-mean and zero cross correlation white Gaussian sequences [Brown et al, 2012].

The extended Kalman filter (EKF) is a nonlinear version of Kalman filtering, which deals with the case described by the nonlinear stochastic differential equations. The EKF is based on consecutive linearization of the nonlinear dynamic and measurement equations about current state estimates. The linearization is typically carried out via Taylor series expansion but interpolation schemes based on numerical differentiation also exist. The linear approximation equations for system and measurement matrices can be obtained in [Zhu et al, 2010]. Since a linear approximation in EKF cannot fully capture the behavior of a nonlinear system. The unscented Kalman filter (UKF) is proposed to improve and simplify the estimation.

The UKF algorithm based on deterministic sampling where, instead of approximating the state and measurement functions of a nonlinear state space model, few deterministic state realizations, so called sigma points, are generated and propagated through the involved nonlinearities. The sample statistics of transformed sigma points are then used in the Kalman filtering equations. The unscented transformation (UT) is the main element of the UKF, where UT is a computation scheme for transforming mean values and covariance matrices that depends on a set of scalar parameters. The fact that UKF variants are sampling based and do not involve Jacobian and Hessian matrices, in contrast to EKF, makes them simple to implement. An extensive survey of these moment computation schemes and their use in Kalman filtering is given in [Julier et al, 2004; Wan et al, 2000]. The underlying idea is to estimate the statistics of the transformed variable from a set of $2n+1$ sigma points, with $n$ being the dimension of the considered random variable. On the basis of the known covariance matrix of the initial random variable, sigma points can be generated deterministically. Unlike the EKF, the UKF does not require the evaluation of the Jacobians of the functions and Hessian matrices, since the gains to be used during the estimation are computed directly from the sigma points. Hence, the UKF represents a possible alternative to the EKF whenever a linearized model is not accurate enough or the Jacobian computation becomes too unwieldy [Jwo et al, 2013]. Figure 1 shows the procedure diagrams to implement the EKF and UKF [Lin, 2017].

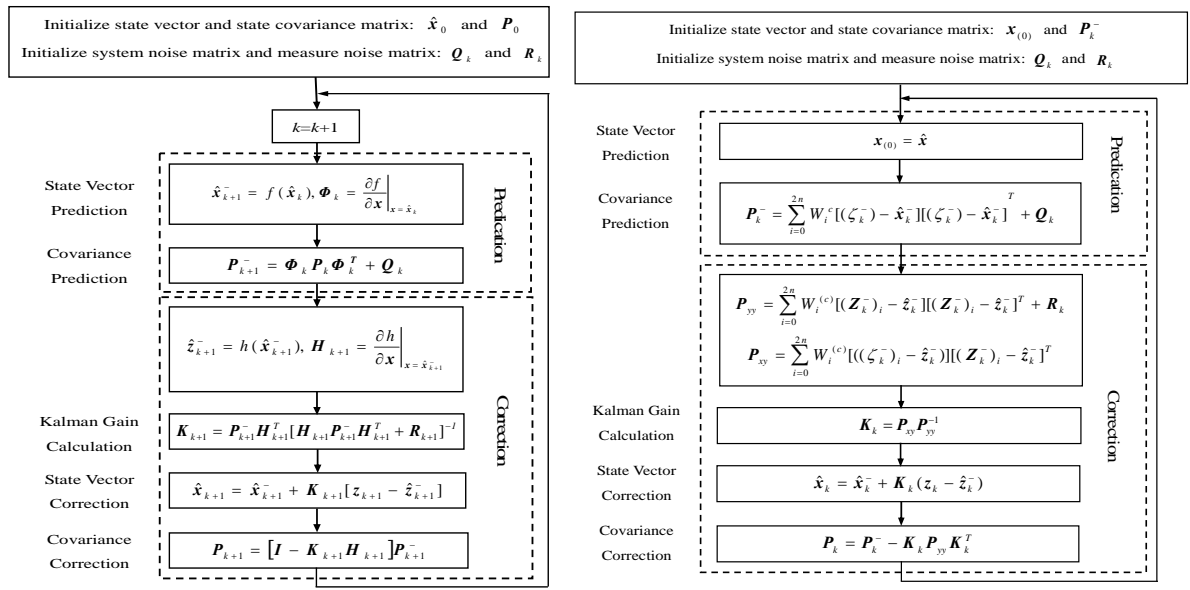

**FIGURE 1. PROCEDURES DIAGRAM FOR THE EKF (LEFT) AND THE UKF (RIGHT)**

## 2.2. Receiver Autonomous Integrity Monitoring (RAIM) and Fault Detection and Exclusion (FDE)

Receiver autonomous integrity monitoring (RAIM) provides integrity monitoring of GPS for aeronautic applications [Lu et al, 2009]. In order for a GPS receiver to perform RAIM function, at least five visible satellites with acceptable geometry must be visible to it. RAIM has various kinds of operations; performing consistency checks between all positions solutions obtained with various subsets of the visible satellites is one of the operations. The receiver provides a vigilant to the pilot if the consistency checks fail. Availability is an important issue and a performance indicator of the RAIM algorithm. Due to geometry and satellite service maintenance, RAIM is not constantly available and antenna of the receiver could have fewer satellites in view occasionally. The RAIM algorithm might be available but with weak capability of detecting a failure when it happens. Therefore, RAIM employed in GPS receivers can be enhanced by applying fault detection and exclusion (FDE) [Liu, 2017]. It uses a minimum of 6 satellites to detect a possible defective satellite and to exclude it from the navigation solution where the navigation function can continue without interruption. The objective of fault detection is to detect the occurrence of a positioning failure [Wu et al, 2017].

**3. LAND MOBILE SATELLITE CHANNEL MODEL**

Land mobile satellite (LMS) channel is an important aspect of GPS system to everyone across the globe due to the services available through it. Therefore, modeling LMS channel in the GPS application services is important to ensure availability, quality of service, and reduce outages on the channel, which in turn will results into improve GPS performance. In this study, the GPS receiver proposed is given simulation over land mobile satellite channel model. The channel model proposed by A. Steingass and A. Lehner (the German Aerospace Centre, DLR) [Steingass et al, 2004; 2005] combining the measured channel impulse response to allow the realistic simulation of the multipath channel by approximating every single reflection. Signal processing over the land mobile satellite (LMS) channel model can be realized independently of the transmitted signal, since the proposed channel model comprises time variant reflectors approaching and lessening in dependency of the azimuth and elevation of the satellite [Lehner, 2007]. Therefore, the LMS channel model is better applied to experiment GPS receiver performance over LMS fading channel situations [Rao et al, 2013]. The LMS channel is mainly impaired by the environment in the direct vicinity of the moving terminal. In an open environment the direct path represented by the line of sight (LOS) transmission of the signal may be totally jammed by concrete substances (e.g., a building, trees, Lampposts) or may be shadowed by flora depending on the density of the vegetation and branches. Also, diffraction, reflection, and scattering of the transmitted signal cause multipath components. In an urban environment the LOS signal is often obstructed so that the first received path is attenuated and possibly delayed with respect to the LOS. Therefore, multipath fading error is the code tracking loop error caused by the LOS direct signals of the satellite and the reflected signals in the vicinity of the receiver.

**4. SIMULATION EXPERIMENT AND RESULTS**

System simulation is given based on the vector delay lock tracking loop GPS processing, where the noise interference portion is set to encounter multi-path fading effects in the signal receiving part. Followed by adapting receiver autonomous integrity monitoring (RAIM) to the GPS tracking loop, and the Kalman filter is performed at the end processing as shown in Fig. 2. The multipath interference simulations are given as Rayleigh Fading and Rician Fading. The estimation filters approved are the extended Kalman filter (EKF) and Unscented Kalman filter (UKF).

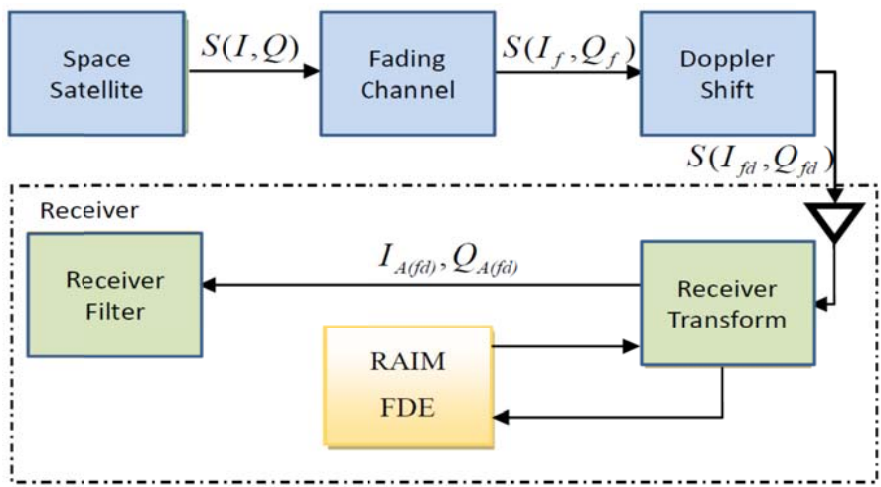

**FIGURE 2. GPS RECEIVER PROCESSING OVER FADING CHANNEL**

The simulation motion trajectory is generated by Inertial Navigation Toolbox 3.0 with Matlab platform. The GPS information is generated by software, Satellite Navigation (SATNAV) Toolbox. Figure 3 shows the three-dimensional vehicle trajectory applied in this study. The receiver autonomous integrity monitoring (RAIM) and fault detection and exclusion (FDE) algorithm are utilized to observe the channel appearing signal interruptions. Simulation parameters are set to positioning every 0.1 seconds for a total 630 seconds period. The error statistics for carrier trajectory about east-west direction, north-south direction, altitude, and root mean square error ($RMS_{error}$) are collected as shown in the corresponding figures.

Figure 4 is an evaluation base of a pre-test condition result for GPS receiver performance without RAIM/FDE algorithms. Applying RAIM algorithm, the GPS navigation filter uses an EKF or a UKF separately. From the simulations, condition 1 is for signal interfering with Rayleigh fading noise with EKF, condition 2 is for signal interfering with Rayleigh fading noise with UKF. Shown in Fig. 5, we may notice that, the UKF GPS receiver performance is worse than the EKF GPS receiver performance. The nonlinear estimation of Kalman filtering parameters are not converged fast enough in UT calculation due to the NLOS situation. Condition 3 is for signal interfering with Rician fading noise with EKF, as well as, condition 4 is for signal interfering with Rician fading noise with UKF, where we have found the UKF GPS receiver performed much better than that in EKF case in Fig. 6. We have also observed from different fading channel conditions, the $RMS_{error}$ for the directions, east-west and north-south, as well as the altitude performs the similar tendency within the same channel. Comparing among the Rayleigh fading and Rician fading, we have found that steady $RMS_{error}$ performance shown at Rician fading channel, since the Rician chanel is include a line-of-sight (LOS) signal path. The worst performance is shown in the Rayleigh distribution channel condition where the multipath noise attenuation between the satellite signal and the receiver is obviously unstable relative to the Rayleigh distribution.

The satellite interruption situation is experimented where we have set three break position. We have noticed that form the channel simulation resulting figures, for Rayleigh distribution fading, at the breaking point, a large tracking loss happened. Since we have applied RAIM algorithm with an EKF or a UKF, from the satellite interruption situation experiment, we may notice that the RAIM algorithm applying can minimize the deep $RMS_{error}$ at the break points over fading channel cases.

## 5. CONCLUSION

Simulations of the satellite signal to GPS receiver over the land mobile satellite multipath channel model are investigated under Rayleigh and Rician fading noise situations. A well-found GPS receiver system provides applicable and truthful positioning information to the users as shown in our simulation results. The tested GPS receiver is supplied with VDLL structure, RAIM-FDE algorithm and the support of Kalman filtering algorithms to improve the system performance over land mobile satellite multipath fading situations. The equipped GPS receiver with RAIM/FDE algorithm is able to prevent signal unlocked situation as shown in our simulations. Both the extended Kalman filter algorithm and/or Unscented Kalman filter applied may improve the system performance. We have found form the simulation results, GPS receiver system with RAIM/FDE supported performs better particularly while satellite interruption situation happened. The navigation accuracy based on the proposed strategy indicates substantial improvement in both positioning accuracy and tracking capability at fading channel conditions, especially in Rician fading environments.

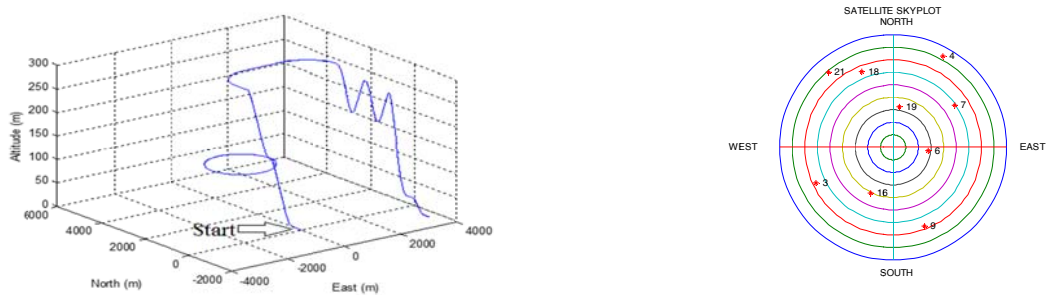

**FIGURE 3. THREE-DIMENSIONAL VEHICLE TRAJECTORY AND SATELLITE DISTRIBUTION LAYOUT APPLIED IN THE EXPERIMENT**

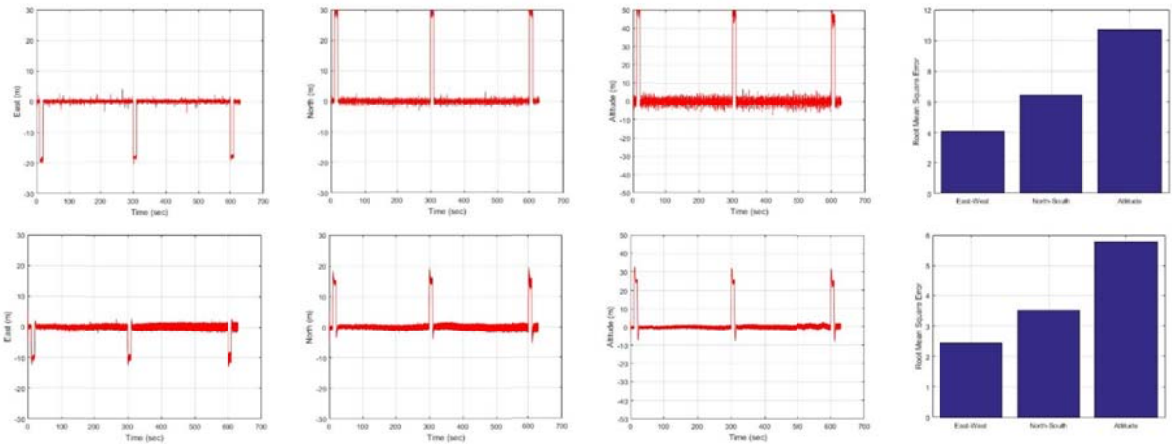

**FIGURE 4. GPS RECEIVER WITHOUT RAIM/FDE, RAYLEIGH FADING (UP: EKF, DOWN: UKF)**

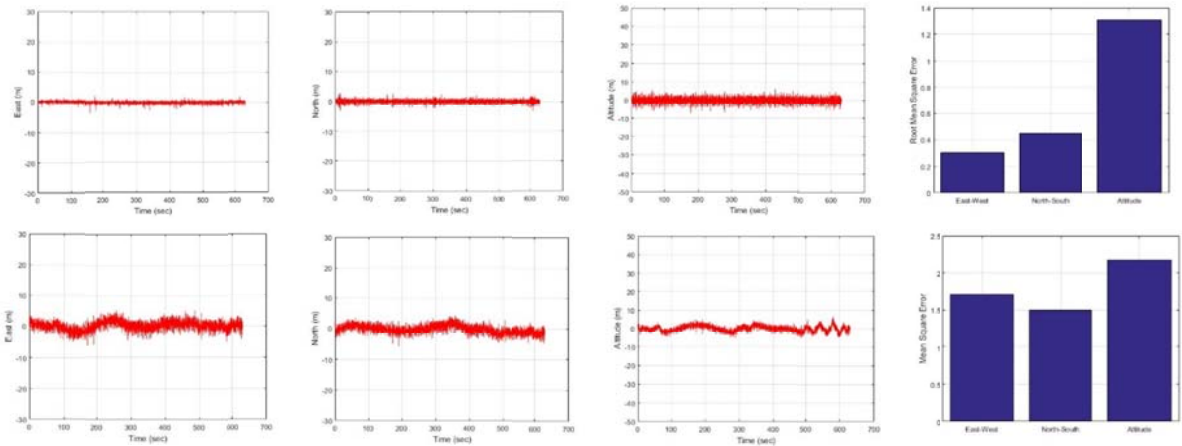

**FIGURE 5. GPS RECEIVER WITH RAIM/FDE, RAYLEIGH FADING (UP: EKF, DOWN: UKF)**

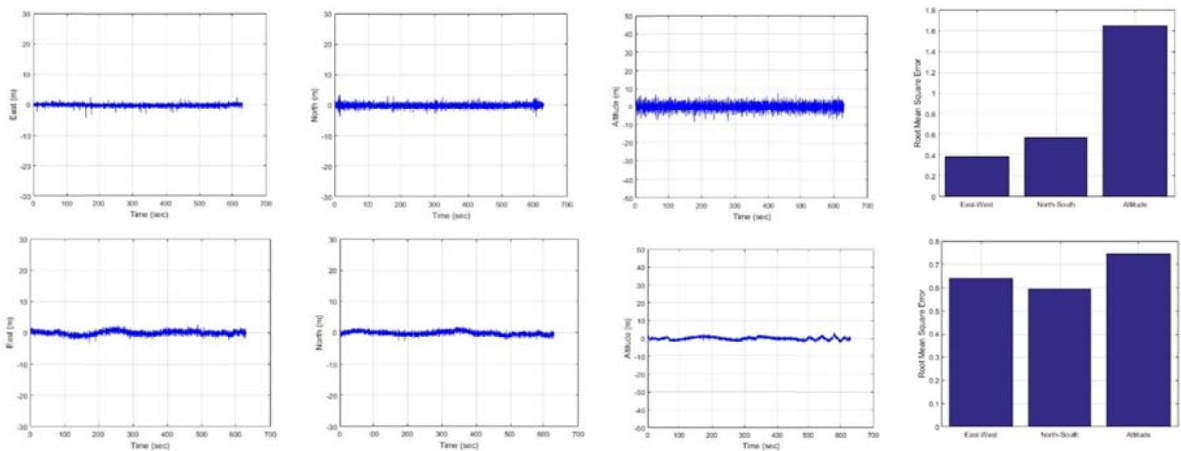

**FIGURE 6. GPS RECEIVER WITH RAIM/FDE, RICIAN FADING (UP: EKF, DOWN: UKF)**

## 6. ACKNOWLEDGEMENT

This work has been supported by the Ministry of Science and Technology, the Republic of China under grant number MOST 106-2119-M-009-004.

**REFERENCES**

Brown, R. G.; Hwang, P. Y. C. 2012. Introduction to Random Signals and Applied Kalman Filtering with Matlab Exercises, 4th ed., Wiley, New York.

Chen, C.-C. 2010. The Vector Tracking Loop Design of GPS Receiver Using the Unscented Kalman Filter. Master's Thesis, National Taiwan Ocean University, Taiwan.

Farrell, J. A.; Barth, M. 1999. The Global Positioning System and Inertial Navigation, McGraw-Hill, New York.

Fontán, F. P.; Vázquez-Castro, M.; Cabado, C. E.*;* Garciá, J. P.; Kubista, E. 2001. "Statistical Modeling of the LMS Channel", *IEEE Transactions on Vehicular Technology*, Vol.50(6), P. 1549-1567.

Jwo, D.-J. 2005. "GPS Navigation Solutions by Analogue Neural Network Least-Squares Processors", *Journal of Navigation*, Vol. 58(1), P. 105-118.

Jwo, D.J., Wang, S.H. 2007. "Adaptive fuzzy strong tracking extended Kalman filtering for GPS navigation", *IEEE Sens. J.*, Vol. 7(5), P. 778–789.

Jwo, D. J.; Yang, C.-F.; Chuang, C.-H.; Lee, T.-Y. 2013. "Performance enhancement for ultra-tight GPS/INS integration using a fuzzy adaptive strong tracking unscented Kalman filter", *Nonlinear Dynamics*, Vol. 73 (1-2), P. 377-395.

Julier, S. J.; Uhlmann, J. K. 2004. "Unscented Filtering and Nonlinear Estimation", *Proceedings of the IEEE*, Vol. 92(3), P. 401–422.

Lashley, M.; et al. 2009. "Performance Analysis of Vector Tracking Algorithms for Weak GPS Signals in High Dynamics", *IEEE Journal of Selected Topics in Signal Processing*, Vol. 3, P. 661-673.

Lashley, M.; Bevly, D. M. 2011. "Comparison in the Performance of the Vector Delay/Frequency Lock Loop and Equivalent Scalar Tracking Loops in Dense Foliage and Urban Canyon", *Proceedings of the International Technical Meeting of The Institute of Navigation*, P. 1786-1803.

Lehner, A. 2007. Multipath Channel Modelling for Satellite Navigation Systems. PhD Thesis, University of Erlangen-Nuremberg, Germany.

Li, H.; Yang, J. 2010. "Analysis and Simulation of Vector Tracking Algorithms for Weak GPS Signal", *International Asia Conference on Informatics in Control, Automation and Robotics*, P. 215-218.

Lin, S.-Y. 2017. Land Mobile Satellite Channel Applied to Autonomous GPS Positioning. Master's Thesis, National Taiwan Ocean University, Taiwan.

Liu, C.-Y. 2017. Integrity Monitoring GPS Vector Tracking Loop over Land Mobile Satellite Channel Analysis. Master's Thesis, National Taiwan Ocean University, Taiwan.

Lu, D. J.; Chen, X. W. 2009. "Algorithm for Global Navigation Satellite System Receiver Autonomous Integrity Monitoring", *Computer Engineering*, Vol.35(11), P. 10-12.

Pany, T.; Eissfeller, B. 2006. "Use of a Vector Delay Lock Loop Receiver for GNSS Signal Power Analysis in Bad Signal Conditions", *IEEE/ION Position, Location, and Navigation Symposium*, P. 893-903.

Peng, S.; Morton, Y.; Di, R. 2012. "A Multiple-Frequency GPS Software Receiver Design Based on A Vector Tracking Loop", *IEEE/ION Position, Location and Navigation Symposium*, P. 495-505.

Rao, G. S. B.; Kumar, G. S.; Kumar, M. N. 2013. "GPS Signal Rician Fading Model for Precise Navigation in Urban Environment", *Indian Journal of Radio & Space Physics*, Vol. 42, P. 192-196.

Spilker Jr., J. J.; Axelrad, P.; Parkinson, B. W.; Enge, P. 1996. Global Positioning System: Theory and Applications Vol. 1, The American Institute of Aeronautics and Astronautics (AIAA), Washington, DC.

Steingass, A.; Lehner, A. 2004. "Measuring the Navigation Multipath Channel-A Statistical Analysis", *Proceedings of the ION GNSS*.

Steingass, A.; Lehner, A. 2005. "A Novel Channel Model for Land Mobile Satellite Navigation", *Proceedings of the ION GNSS*, P. 2132-2138.

Wan, E. A.; van der Merwe, R. 2000. "The Unscented Kalman Filter for Nonlinear Estimation", *Proceedings of IEEE Symposium AS-SPCC*.

Wu, J.-C.; Jwo, D.-J.; Chien, C.-Y. 2017. "Adaptive Vector Delay Lock Tracking Loops GPS Signal over Multipath Fading Channels", *ACM Conference Proceeding-WCNA2017*, P. 194-198.

Zhu, Z.; Cheng, Z.; Tang, G.; Li, S.; Huang, F. 2010. "EKF Based Vector Delay Lock Loop Algorithm for GPS Signal Tracking", *International Conference on Computer Design and Applications*, P. 352-356.